\documentclass[%
 aip,
 jap,%
 amsmath,amssymb,
 reprint,%
]{revtex4-1}

\usepackage{graphicx}
\usepackage{dcolumn}
\usepackage{bm}

\begin{document}


\title{Suppressing nano-scale stick-slip motion by feedback}

\author{Jing Zhang}
\email{jing-zhang@mail.tsinghua.edu.cn}
\affiliation{Department of Automation, Tsinghua University,
Beijing 100084, P. R. China}
\affiliation{Center for Quantum
Information Science and Technology, Tsinghua National Laboratory
for Information Science and Technology, Beijing 100084, P. R.
China}
\author{Re-Bing Wu}
\affiliation{Department of Automation, Tsinghua University,
Beijing 100084, P. R. China}
\affiliation{Center for Quantum
Information Science and Technology, Tsinghua National Laboratory
for Information Science and Technology, Beijing 100084, P. R.
China}
\author{Lei Miao}
\affiliation{State Key Laboratory of Robotics, Shenyang Institute
of Automation, Chinese Academy of Sciences, Beijing, 110016,
China}
\author{Ning Xi}
\affiliation{Department of Electrical and Computer Engineering,
Michigan State University, East Lansing, MI 48824, USA}
\author{Chun-Wen Li}
\affiliation{Department of Automation, Tsinghua University,
Beijing 100084, P. R. China}
\affiliation{Center for Quantum
Information Science and Technology, Tsinghua National Laboratory
for Information Science and Technology, Beijing 100084, P. R.
China}
\author{Yue-Chao Wang}
\affiliation{State Key Laboratory of Robotics, Shenyang Institute
of Automation, Chinese Academy of Sciences, Beijing, 110016,
China}
\author{Tzyh-Jong Tarn}
\affiliation{Center for Quantum Information Science and
Technology, Tsinghua National Laboratory for Information Science
and Technology, Beijing 100084, P. R. China}
\affiliation{Department of Electrical and Systems Engineering,
Washington University, St.~Louis, MO 63130, USA}

\date{\today}

\begin{abstract}
When a micro cantilever with a nano-scale tip is manipulated on a
substrate with atomic-scale roughness, the periodic lateral
frictional force and stochastic fluctuations may induce stick-slip
motion of the cantilever tip, which greatly decreases the
precision of the nano manipulation. This unwanted motion cannot be
reduced by open-loop control especially when there exist parameter
uncertainties in the system model, and thus needs to introduce
feedback control. However, real-time feedback cannot be realized
by the existing virtual reality virtual feedback techniques based
on the position sensing capacity of the atomic force microscopy
(AFM). To solve this problem, we propose a new method to design
real-time feedback control based on the force sensing approach to
compensate for the disturbances and thus reduce the stick-slip
motion of the cantilever tip. Theoretical analysis and numerical
simulations show that the controlled motion of the cantilever tip
tracks the desired trajectory with much higher precision. Further
investigation shows that our proposal is robust under various
parameter uncertainties. Our study opens up new perspectives of
real-time nano manipulation.
\end{abstract}

\maketitle

%

\section{Introduction}\label{s1}
One of the central problems in nano science and
technology~\cite{Feynman} is the realization of high-precision
nano manipulation, e.g., pushing, pulling, rotating, rolling, and
cutting nano-scale objects. In the widely applied AFM
experiments~\cite{Binnig1,Binnig2,Junno,Martin,Guthold,Sitti1,Sitti2,XiNing1,XiNing2,TianXiaoJun,LiuLianQing,ZhangYu},
a large amount of frontier progresses have been achieved about
nano manipulation. However, the theoretical analysis for such
system is mainly focused on the static force analysis within a
Newton-mechanical
framework~\cite{Junno,Martin,Guthold,Sitti1,Sitti2,XiNing1,XiNing2,TianXiaoJun,LiuLianQing,ZhangYu}
that describes the macroscopic system. Recent experiments show
that the force analysis in the nano scale is not the same as that
at the macroscopic scale. Typically, nano-scale
friction~\cite{PerssonBook,PerssonRMP} is logarithmically
dependent on the velocity~\cite{Gnecco,Riedo,Tambe}, which is
quite different from the macroscopic sliding friction (the
friction is independent on the velocity) and the traditional
viscous-type friction (the friction is linearly proportional to
the velocity). Additionally, the atomic scale periodic structure
on the substrate has to be seriously considered in the friction
analysis that is usually done with the continuous mechanics in
macroscopic system~\cite{Tshiprut,Socoliuc1}.

The periodic lateral force induced by the atomic periodic
structure on the surface of the substrate may lead to stick-slip
motions~\cite{Heslot,Conley,Braiman1} of the cantilever tip of the
AFM, which deteriorates the precision of the nano manipulation. In
the literature~\cite{Tshiprut,Socoliuc1}, the cantilever of the
AFM is usually modelled as a soft spring, and the effective nano
friction force is described by a sinusoidal lateral force.
Temperature-dependent white noises are introduced to represent the
stochastic fluctuations in the lateral force induced by the
thermal motion of the substrate atoms. Such a dynamical model
predicts a motion transition of the cantilever tip of the AFM from
the continuous sliding to a stick-slip mode by varying the ratio
between the amplitude of the sinusoidal lateral force and the
stiffness of the cantilever tip.

To improve the precision in nano manipulation, many
strategies~\cite{Braiman2,Socoliuc2,Guerra,Capozza,Lizuka,GuoY1,GuoY2}
have been proposed to control the motions of the nano objects
under nano frictions. For example, in the
literature~\cite{Braiman2}, the authors designed a
non-Lipschitzion control function under which one can push the
nano sample to asymptotically track the target velocity. This
proposal is efficient and robust, but the natural fluctuation
cannot be removed. To overcome this difficulty, more complex
control function was designed based on the Lyapunov theory in
Refs.~\cite{GuoY1,GuoY2}. However, the control designs in these
proposals require the knowledge of the exact position of the
sample during the nano manipulation. Such schemes are uneasy to be
realized with the present experimental techniques, e.g., the
haptic sensing and the virtual reality visual feedback
techniques~\cite{XiNing1,XiNing2,Sitti3}, for the inability of
simultaneous position sensing and manipulation processing by the
AFM. In fact, in these techniques, one has to stop the nano
manipulation process and scan the surface of the substrate by the
cantilever tip of the AFM to position the sample on the substrate,
after which the next step of nano manipulation can go on. To solve
this problem, in this paper, we propose a feedback control
strategy based on the real-time signal sampled from the force
sensor of the AFM, which can be conditionally done without pending
the nano manipulation. The signal is used to estimate the position
of the cantilever tip of the AFM, with which we can design
feedback control to reduce the stick-slip motion of the cantilever
tip and thus improve the manipulation precision.

The paper is organized as follows. Section~\ref{s2} describes the
dynamical model, following which the stick-slip motions of the
cantilever tip under open-loop control are presented in
Sec.~\ref{s3}. Section~\ref{s4} is devoted to the design of the
real-time feedback control to reduce the stick-slip motion of the
cantilever tip. Section~\ref{s5} discusses the robustness of our
method against various parameter uncertainties. Conclusions and
forecast of the future work are given in Sec.~\ref{s6}.

\section{Modelling of the one-dimensional nano manipulation
system}\label{s2}

We first present the model used to describe the one-dimensional
nano manipulations such as pushing a nano sample or etching the
surface of the substrate to draw desired pattern. In this model,
the cantilever of the AFM is taken as a spring with an effective
stiffness $k_c$. Thus, according to the Hooke's law, the lateral
force imposed by the cantilever tip is:
\begin{equation}\label{Pushing force executed by the AFM}
F_p=k_c(u-x),
\end{equation}
where $x$ and $u$ are the relative positions of the cantilever tip
with and without deformation on the platform of the nano
manipulation, respectively (see Fig.~\ref{Schematic diagram of the
one dimensional nano manipulation system}).

In our method, $u$ is the control parameter to be designed. The
mechanism of control is shown in Fig.~\ref{Schematic diagram of
the one dimensional nano manipulation system}(a). As shown in
Fig.~\ref{Schematic diagram of the one dimensional nano
manipulation system}(a), in the designed nano manipulation system,
the cantilever of the AFM is fixed, while the platform of the nano
manipulation moves. In such a system, the motion of the platform
of the nano manipulation is controlled by a piezoelectric
transducer (PZT). {\it By adjusting the voltage $V_{\rm PZT}$
added on the PZT via a control circuit, the deformation of the PZT
can be controlled, by which the relative position $u$ between the
cantilever of the AFM and the platform is tunable.} Typically, the
functional relationship between the deformation of the PZT and the
voltage $V_{\rm PZT}$ shows hysteresis, creep, and structural
vibration behaviors and thus is nonlinear. However, such a
nonlinear characteristic response of PZT can be compensated by
auxiliary control devices. Our recent theoretical and experimental
study~\cite{HuangFQ} shows that a linear functional relationship
between $V_{\rm PZT}$ and $u$ can be obtained by introducing
optimal design of feedforward controller by the Prandtl-Ishlinskii
model~\cite{Brokate}. Based on this study, in order to simplify
our discussions, we do not go into details about the relationship
between $V_{\rm PZT}$ and $u$, {\it but simply take the
deformation of the PZT represented by $u$ as the control
parameter}. Additionally, in the moving frame of the platform, the
control system shown in Fig.~\ref{Schematic diagram of the one
dimensional nano manipulation system}(a) is equivalent to that
shown in Fig.~\ref{Schematic diagram of the one dimensional nano
manipulation system}(b), in which the platform is fixed and the
cantilever moves.

\begin{figure}[h]
\centerline{\includegraphics[bb=83 340 503 789,width=4.3 cm,
clip]{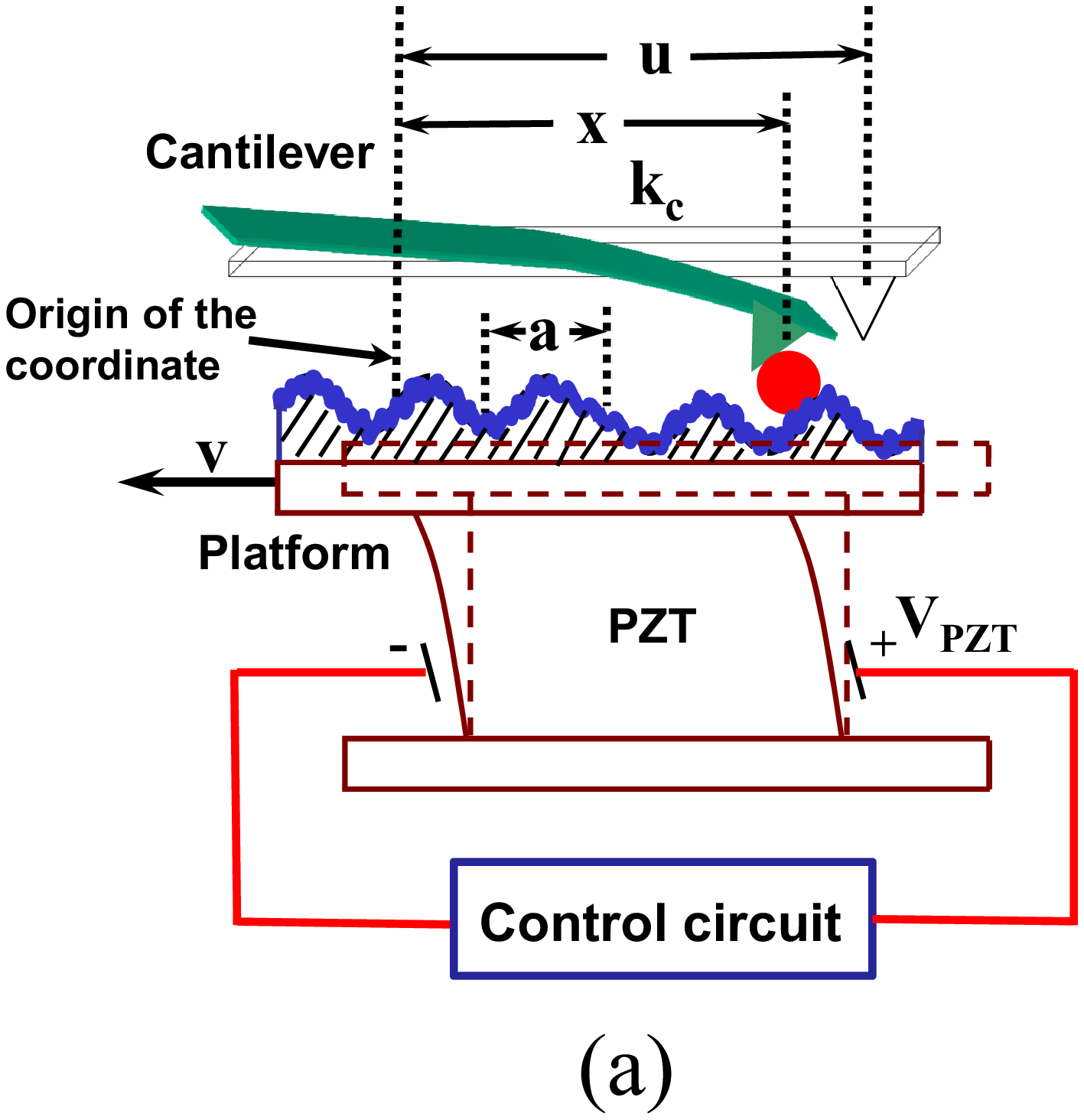}
\includegraphics[bb=98 338 507 758,width=4.3 cm,
clip]{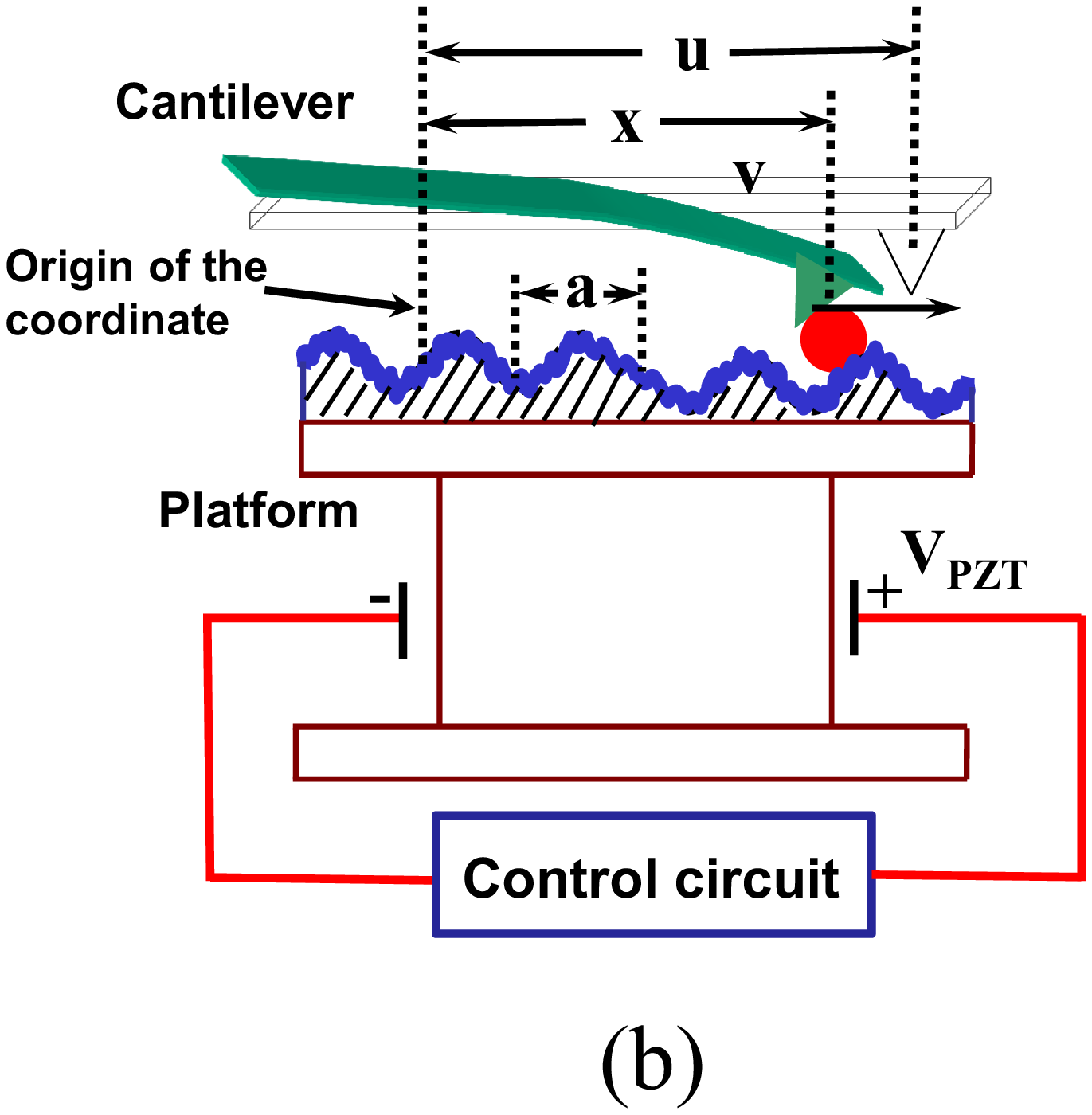}}
\caption{\scriptsize (color online) Schematic diagrams of the
controlled nano manipulation system: (a) the control system in
experiments in which the cantilever of the AFM is fixed and the
platform moves; (b) the equivalent control system in which the
platform is fixed while the cantilever moves. The two control
systems (a) and (b) are equivalent if we fix the origin of the
coordinate with the moving platform. ``PZT" denotes the
piezoelectric transducer. The deformation of the PZT is controlled
by the voltage $V_{\rm PZT}$ added on the PZT. Thus, the relative
position $u$ can be controlled as what we want by a control
circuit.}\label{Schematic diagram of the one dimensional nano
manipulation system}
\end{figure}

When the cantilever tip moves on the substrate, the nano friction
may occur on the contact area. In recent
experiments~\cite{Gnecco}, the average frictional force is
observed to be logarithmically dependent on the velocity of the
cantilever tip $v=dx/dt$, i.e.,
\begin{equation}\label{Average frictional force}
\bar{F}_f=F_{f_0}+F_{f_1}\ln\left(\frac{v}{v_1}\right),
\end{equation}
where $F_{f_0}$, $F_{f_1}$, $v_1$ are constant parameters. There
also exists periodic lateral force induced by the surface
potential of the substrate, and stochastic noises induced by the
thermal motions of the substrate atoms. For simplicity, we only
consider the fundamental-frequency component of the periodic
lateral force, and omit the correlation effects of the thermal
motions of the substrate atoms. Additionally, to simplify our
discussions, the origin of coordinate is chosen such that the
periodic lateral force is zero at the origin. Thus, the resulting
modification of the lateral force can be expressed as:
\begin{equation}\label{Flutuations in the lateral force}
\delta F_f=F_{f_2}\sin\left(2\pi\frac{x}{a}\right)-\xi(t),
\end{equation}
where $F_{f_2}$ is the amplitude of the periodic lateral force;
$a$ is the lattice constant of the substrate; and $\xi(t)$ is a
white noise such that $E(\xi(t))=0$,
$E(\xi(t)\xi(t'))=D_{\xi}\delta(t-t')$, with $D_{\xi}$ being the
strength. $E\left(\cdot\right)$ denotes the ensemble average over
the white noise.

From Eqs.~(\ref{Pushing force executed by the AFM}), (\ref{Average
frictional force}), and (\ref{Flutuations in the lateral force}),
the equations of the system can be expressed in the Ito notation
as:
\begin{eqnarray}\label{Dynamical equations of the sample without defection}
dx & = & v dt,\nonumber\\
m dv & = & k_c(u-x)dt+\left[-F_{f_0}-F_{f_1}\ln\left(\frac{v}{v_1}\right)\right.\nonumber\\
&&\left.-F_{f_2}\sin\left(2\pi\frac{x}{a}\right)\right]dt+\sqrt{D_{\xi}}dW,
\end{eqnarray}
where $m$ is the effective mass of the cantilever tip; and
$dW=W\left(t+dt\right)-W\left(t\right)$ is the increment of the
Wiener process $W\left(t\right)=\int_0^t \xi\left(\tau\right)
d\tau$ satisfying:
\begin{equation}\label{Wiener increments}
E(dW)=0,\quad (dW)^2=dt.
\end{equation}

\section{Stick-slip motion under open-loop control}\label{s3}
In order to implement nano manipulation by the cantilever tip of
the AFM with a constant velocity $v^*$, i.e., to control the
position of the cantilever tip such that $x=v^* t$, a simple
strategy is to set $u=v^* t$. However, this would be ineffective
due to the existence of the periodic lateral force and the
stochastic fluctuations especially when the roughness of the
substrate is relatively large compared with the stiffness of the
cantilever of the AFM. In such case, existing
studies~\cite{Tshiprut,Socoliuc1} have predicted stick-slip
motions of the cantilever tip, which greatly reduce the precision
of the nano manipulation. The lateral pushing force imposed by the
cantilever tip oscillates under the stick-slip motion, which can
be so large that the fragile sample and substrate are damaged, or
the cantilever tip of the AFM slide over them.

The drawbacks of the open-loop constant control can be seen from
the following numerical examples. The system parameters are chosen
as~\cite{Tshiprut}:
\begin{eqnarray}\label{System parameters}
&F_{f_0}=10\,\,{\rm nN},\quad F_{f_1}=1\,\,{\rm nN},\quad a=0.25\,\,{\rm nm},&\nonumber\\
&v_1=1\,\,{\rm
nm/s},\quad F_{f_2}=\,\,{\rm either}\,\,0.25\,\,{\rm nN}\,\,{\rm or}\,\,15\,\,{\rm nN},&\nonumber\\
&v^*=3\,\,{\rm nm/s}, \quad \sqrt{D_{\xi}}=0.1\,\,{\rm nN\cdot
s^{-1/2}},\nonumber\\
&k_c=1\,\,{\rm N/m},\quad m=5\times10^{-11}\,\,{\rm kg},&
\end{eqnarray}
under which the simulation results of the position $x$, velocity
$v$, and the lateral force $F_p$ imposed by the cantilever tip are
shown in Fig.~\ref{Plot of the motion of the sample pushed by
constant control}.

\begin{figure}[h]
\centerline{\includegraphics[bb=88 268 506 588,width=4 cm,
clip]{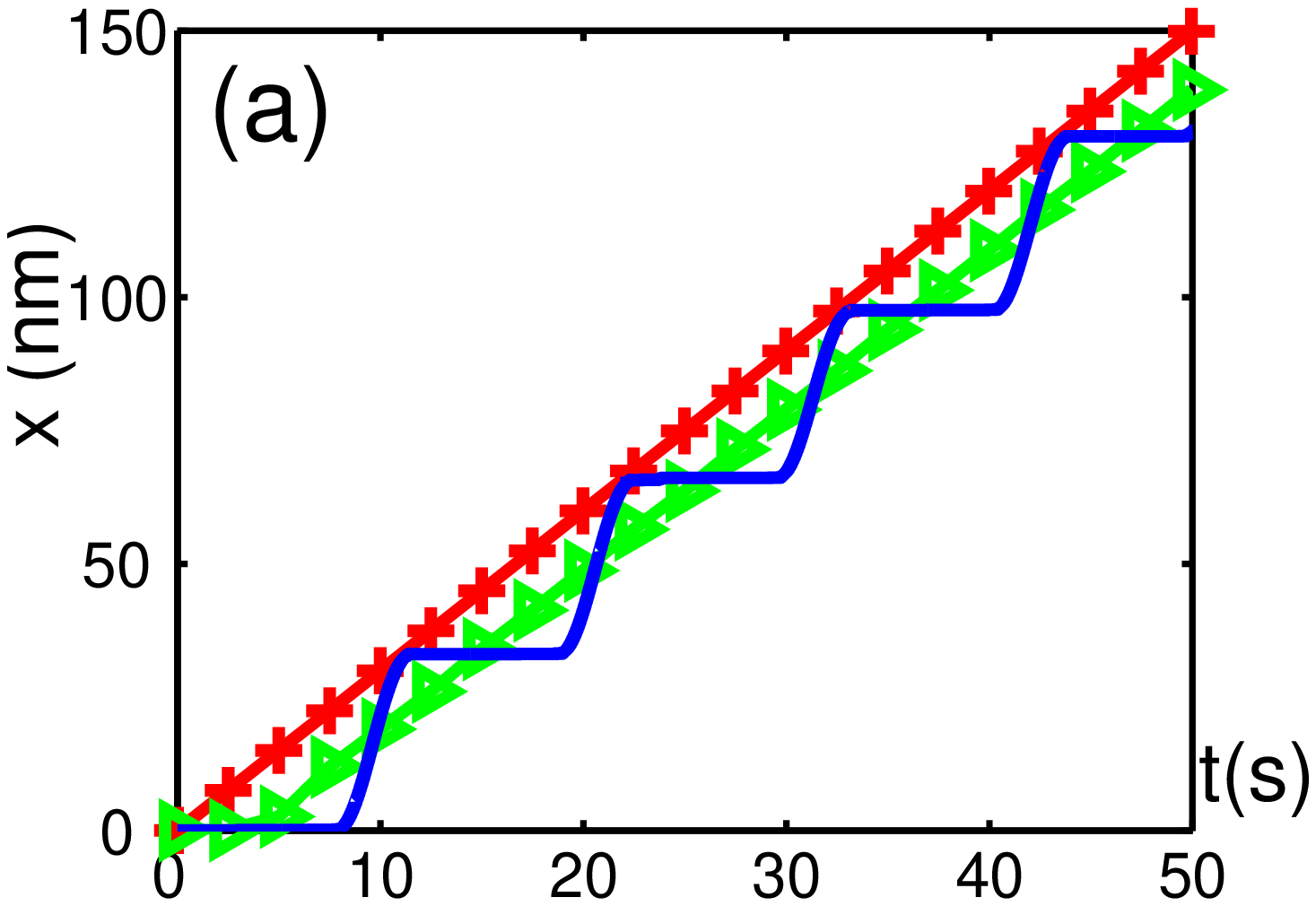}
\includegraphics[bb=96 263 503 569,width=4 cm, clip]{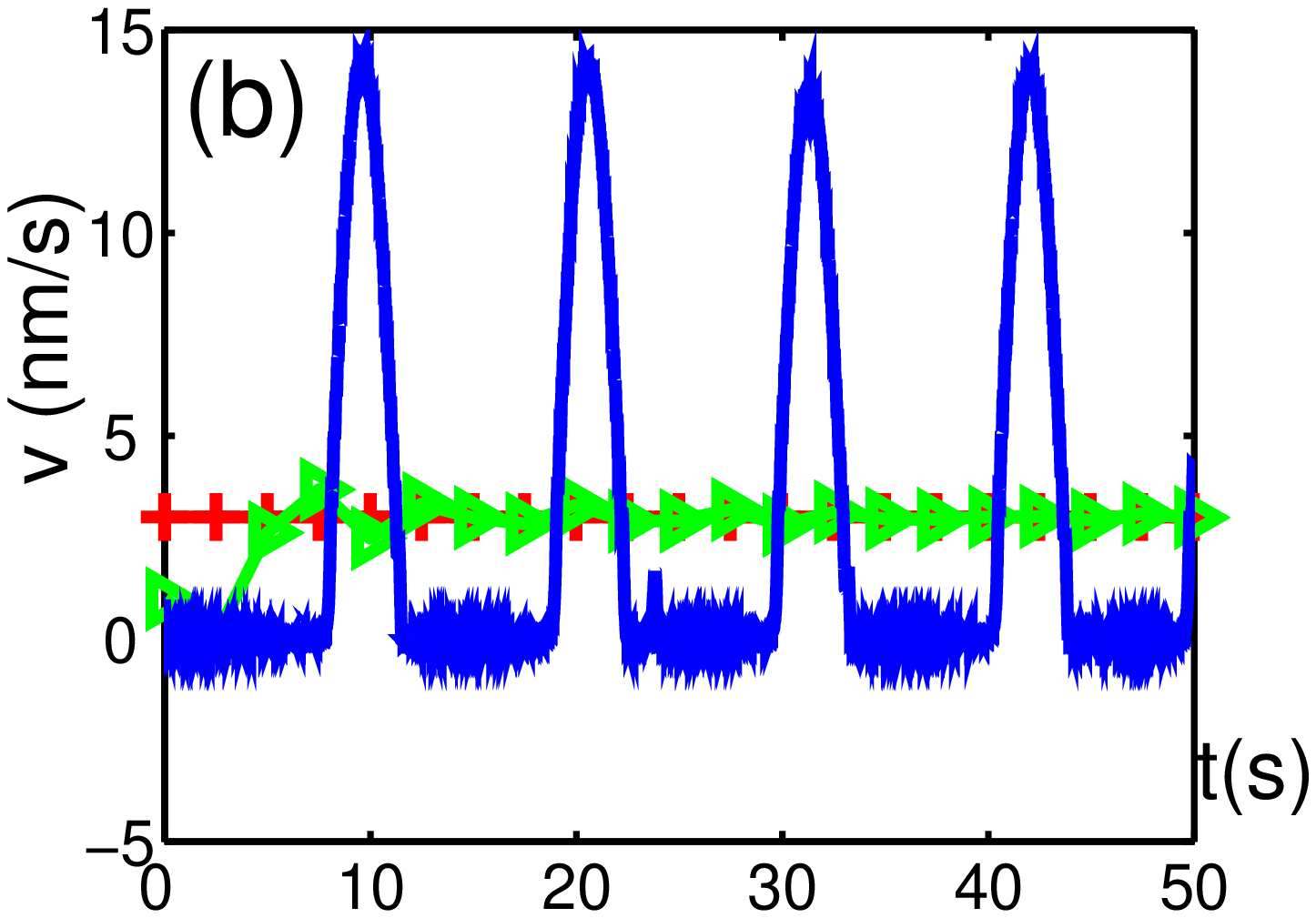}}
\centerline{\includegraphics[bb=88 263 503 579,width=4 cm,
clip]{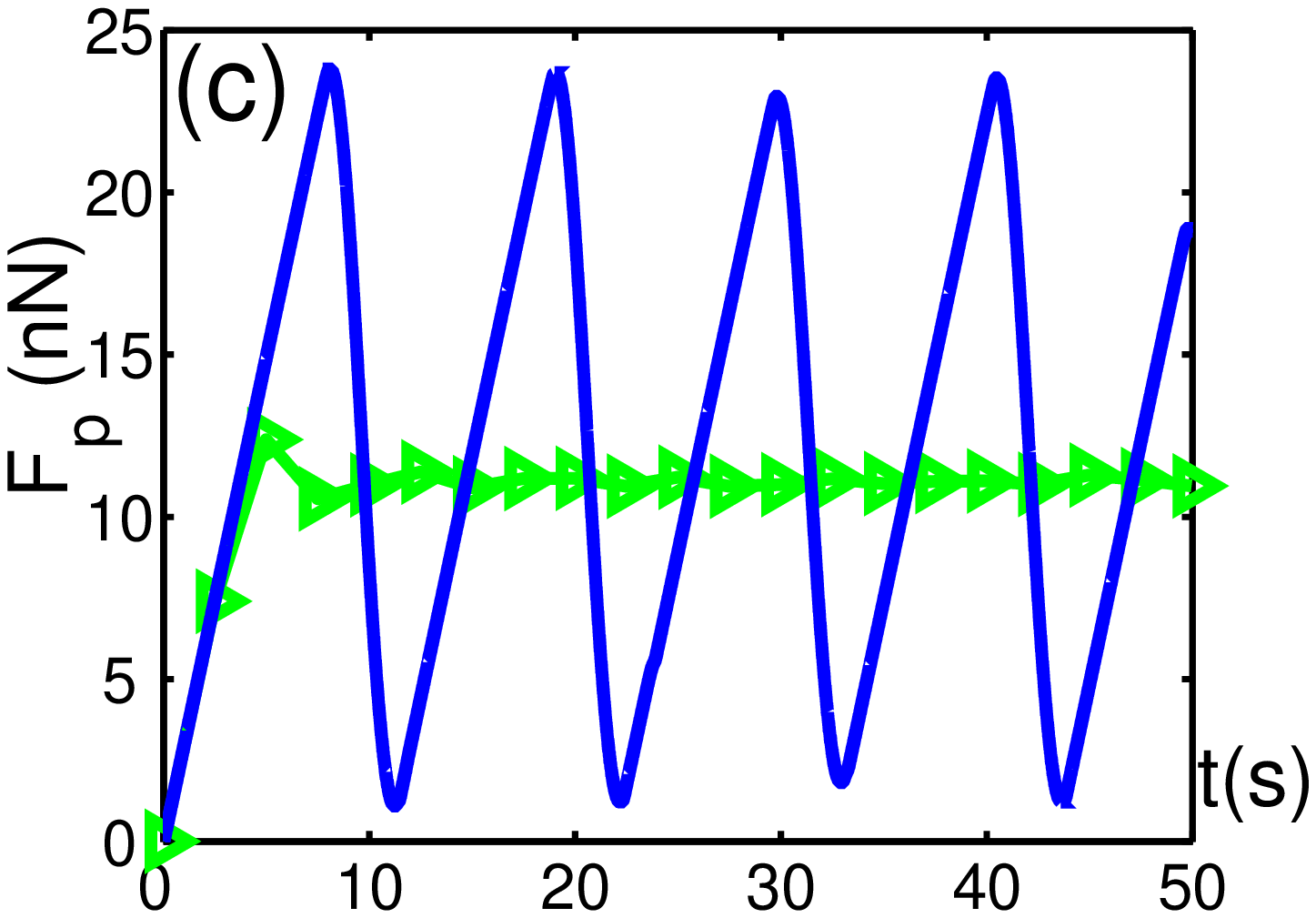}}
\caption{\scriptsize (color online) Plots of (a) the positions,
(b) the velocities of the cantilever tip, and (c) the pushing
forces imposed by the cantilever tip. The red curves with plus
signs denote the ideal trajectories for which the cantilever tip
moves with constant velocity $v=v^*=3$ nm/s; the green triangle
(blue solid) curves represent the controlled trajectories under
open-loop control $x=v^* t$, with the amplitudes of the periodic
lateral force $F_{f_2}$ as $0.25$ nN ($15$ nN).}\label{Plot of the
motion of the sample pushed by constant control}
\end{figure}

As shown in Fig.~\ref{Plot of the motion of the sample pushed by
constant control}, when the amplitude of the periodic lateral
force $F_{f_2}$ is relatively small such that $F_{f_2}/a$ is
comparable with $k_c$ (the case with $F_{f_2}=0.25$ nN), which is
the case when the cantilever tip moves on an incommensurate
substrate~\cite{PerssonBook}, there exists no stick-slip motion
and the cantilever tip moves with a constant velocity after a
transient process (see the green triangle curves). However, when
the amplitude of the periodic lateral force is relatively large
such that $F_{f_2}/a\gg k_c$ (the case with $F_{f_2}=15$ nN),
which may be valid when the cantilever tip moves on a commensurate
substrate, stick-slip motions can be observed (see the blue solid
curves). Such stick-slip motions can be explained by the elastic
instability predicted by the Frenkel-Kontorova
model~\cite{PerssonBook}. In this case, the cantilever tip moves
only when the pushing force $F_p$ exceeds the critical value,
i.e., the peak value of the sawtoothlike trajectory shown in
Fig.~\ref{Plot of the motion of the sample pushed by constant
control}(c). It is also shown in Fig.~\ref{Plot of the motion of
the sample pushed by constant control}(a) that no matter whether
the stick-slip motion occurs or not, there exist tracking errors
between the controlled trajectories (the green triangle curve and
the blue solid curve) and the ideal trajectory (the red curve with
plus signs), which need to be compensated to improve the precision
of the nano manipulation.

\section{Reduction of the stick-slip motion by real-time feedback control}\label{s4}
In this section, we introduce feedback control to reduce this
unexpected stick-slip motion under open-loop control. The object
of our method is to monitor the deformation of the cantilever and
thus make the cantilever tip move with a given constant velocity.

To acquire feedback signals, there are two available sensing
methods supported by the AFM-based nano manipulation systems:
position sensing and force sensing. Position sensing is one of the
basic functions of the AFM, by which one can easily obtain the
tomography of a surface with nano-scale roughness. However,
real-time position sensing is unavailable during the nano
manipulation. Another sensing capacity of the AFM is the force
sensing, which can be done by detecting the deformation of the
cantilever of the AFM. In contrast with position sensing, force
sensing can be done during the nano manipulation, which makes the
real-time feedback control possible. {\it The main idea of our
feedback control method is to estimate the position of the
cantilever tip by force sensing to adjust the control parameter
$u$, and further control the motion of the tip.} The control
process thus can be divided into two steps, i.e., a filtering and
estimation step and a feedback control step, which will be
specified below.

\subsection{Position estimation by force sensing}\label{s41}
By the force sensing capacity of the AFM, the pushing force
measured can be expressed as:
\begin{equation}\label{Measurement equation}
F_p^m=k_c\left(u-x\right)+\eta(t),
\end{equation}
where $\eta\left(t\right)$ is a white noise induced by the
measurement apparatus. $\eta\left(t\right)$ satisfies
\begin{equation}\label{Ensemble average and covariance of eta(t)}
E\left(\eta\left(t\right)\right)=0,\quad
E\left(\eta\left(t\right)\eta\left(t'\right)\right)=D_{\eta}\delta\left(t-t'\right),
\end{equation}
where $D_{\eta}$ is the strength of the noise
$\eta\left(t\right)$.

To reduce the measurement-induced disturbance of
$\eta\left(t\right)$ in feedback control, we filter the measured
signal $F_p^m$ by a low-pass filter over a time window $[t-T,t]$.
The output signal can be expressed as (see, e.g.,
Refs.~\cite{SarovarM,Liuzhuo}):
\begin{equation}\label{Estimation of the measured pushing force}
\hat{F}_p^m\left(t\right)=\frac{1}{T}\int_{t-T}^t
e^{-\gamma_{ft}\left(t-\tau\right)}F_p^m\left(\tau\right)d\tau,
\end{equation}
where $\gamma_{ft}$ is the damping rate of the low-pass filter.
Then, the position of the tip $x$ can be estimated from
$\hat{F}_p^m$ by:
\begin{equation}\label{Estimations of the position}
\hat{x}\left(t\right)=u(t)-\frac{1}{k_c}\hat{F}_p^m\left(t\right).
\end{equation}
Under the filtering condition
\begin{equation}\label{Filtering condition}
\gamma_{ft}\ll1/T,
\end{equation}
it can be verified that
\begin{equation}\label{Exponential convergence of the position estimation}
x\left(t\right)-\hat{x}\left(t\right)=e^{-\gamma_{ft}t}\left(x_0-\hat{x}_0\right),
\end{equation}
where $x_0$ and $\hat{x}_0$ are the initial states of
$x\left(t\right)$ and $\hat{x}\left(t\right)$ respectively (see
the derivation in Appendix~\ref{Derivation of the exponential
convergence of the position estimation}). Thus, the estimated
position $\hat{x}\left(t\right)$ exponentially converges to the
position of the tip $x\left(t\right)$ with the convergence rate
$\gamma_{ft}$, and thus $\hat{x}\left(t\right)$ can be taken as a
good estimation of $x\left(t\right)$ when $t\gg 1/\gamma_{ft}$.

\subsection{Feedback control design based on the estimated position}
Based on the estimated position $\hat{x}$ given in
Eq.~(\ref{Estimations of the position}), we can design the
position-based feedback control $u\left(\hat{x}\right)$ to reduce
the stick-slip motion of the cantilever tip, which is given as
follows:
\begin{eqnarray}\label{Real-time feedback control}
u\left(\hat{x}\right)&=&\frac{1}{k_c}\left[k_c
x^*+F_{f_0}+F_{f_1}\ln\left(\frac{v^*}{v_1}\right)+F_{f_2}\sin\left(2\pi\frac{\hat{x}}{a}\right)\right.\nonumber\\
&&\left.-k_x\left(\hat{x}-x^*\right)-k_I\int_0^t\left(\hat{x}\left(\tau\right)-x^*\right)d\tau\right],
\end{eqnarray}
where $x^*=v^*t$ is the desired motion of the tip; and
$k_x,\,k_I>0$ are the control parameters to be determined. The
term
\begin{eqnarray*}
k_c x^*+F_{f_0}+F_{f_1}{\rm
ln}\left(v^*/v_1\right)+F_{f_2}\sin\left(2\pi\hat{x}/a\right)
\end{eqnarray*}
in the control given by Eq.~(\ref{Real-time feedback control}) is
introduced to compensate the mean and the periodic friction
forces. The proportional and integral feedback control terms
$k_x\left(\hat{x}-x^*\right)$ and
$k_I\int_0^t\left(\hat{x}-x^*\right)d\tau$ are introduced to speed
up the convergence of the system dynamics to the stationary motion
and reduce the static error.

\begin{figure}[h]
\centerline{\includegraphics[width=8 cm]{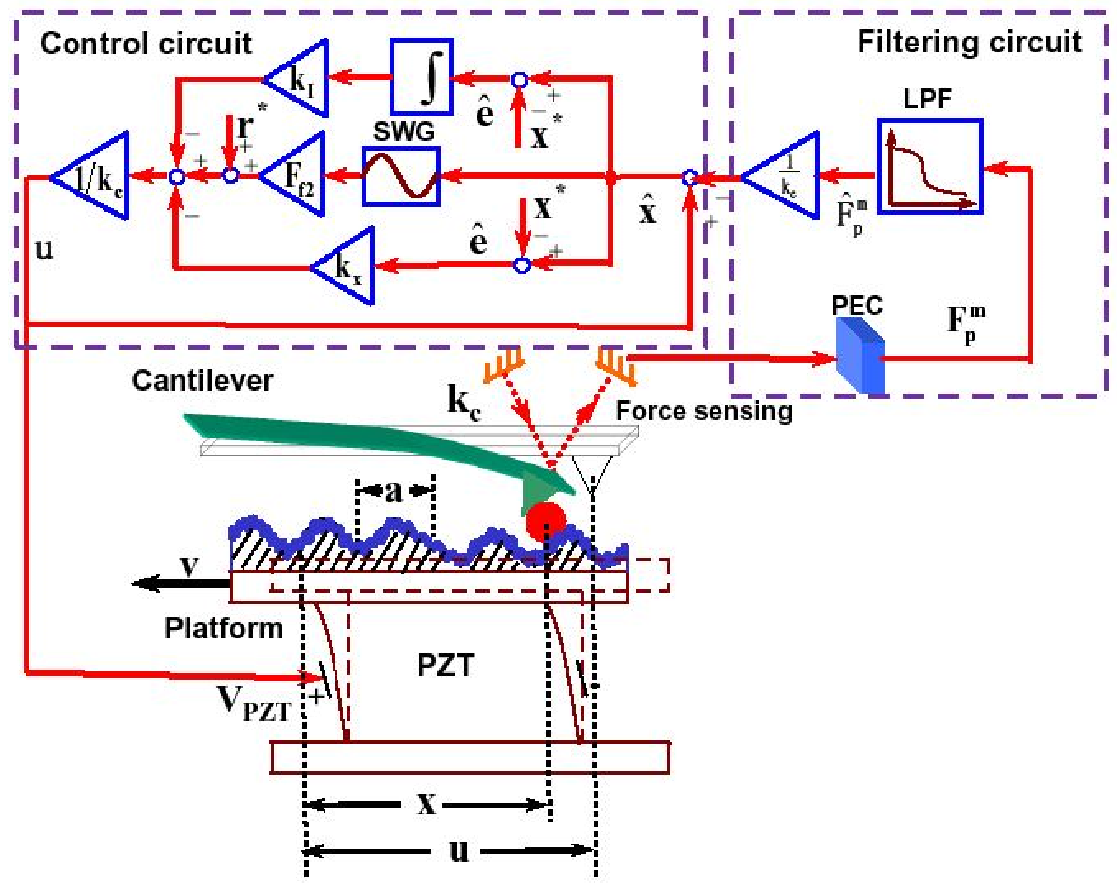}}
\caption{\scriptsize (color online) Schematic diagram of the
feedback control loop, where $r^*=k_c x^*+F_{f_0}+F_{f_1}{\rm
ln}\left(v^*/v_1\right)$ and $\hat{e}=\hat{x}-x^*$; ``PZT",
``PEC", ``LPF", and ``SWG" are the piezoelectric transducer, the
photonelectric convertor, the low-pass filter, and the sine-wave
generator respectively. The feedback control circuit is divided
into two parts, i.e., a filtering circuit and a control circuit
The designed feedback control $u$ in Eq.~(\ref{Real-time feedback
control}) can be used to find the voltage $V_{\rm PZT}=V_{\rm
PZT}\left(u\right)$ added on the PZT.}\label{Schematic diagram of
the feedback control loop}
\end{figure}

The schematic diagram of the feedback control proposed is given in
Fig.~\ref{Schematic diagram of the feedback control loop}. The
deformation of the cantilever tip of the AFM is detected by an
optical refracting system, and then converted into electric
signals by a photonelectric convertor. The output signal masked by
the measurement noise $\eta\left(t\right)$ is fed into a low-pass
filter followed by a control circuit to generate the feedback
control signal. The feedback electric signal is used to control
the motion of the platform by adjusting the voltage $V_{\rm PZT}$
added on the piezoelectric transducer connected to the platform.

Under the filtering condition~(\ref{Filtering condition}), the
weak noise condition
\begin{equation}\label{Tiny fluctuation condition}
D_{\xi}\ll2F_{f_1}\left(k_c+k_x\right)a^2/v^*,\,2F_{f_1}mv^*,
\end{equation}
and choosing the control parameters $k_I$ and $k_x$ such that
\begin{equation}\label{kI without uncertainties}
k_I<F_{f_1}\left(k_c+k_x\right)/mv^*,
\end{equation}
the controlled trajectory $x\left(t\right)$ tracks the desired
trajectory $x^*=v^*t$ in average (see Appendix~\ref{Derivations of
the main results in Sec.IV}), i.e.,
\begin{equation}\label{Position tracking}
\lim_{t\rightarrow\infty}\left(E\left(x\left(t\right)\right)-x^*\right)=0,
\end{equation}
where $E\left(\cdot\right)$ denotes the ensemble average of the
stochastic signal. This result indicates that the stick-slip
motion of the cantilever tip can be efficiently suppressed by the
designed feedback control.

To evaluate the magnitude of the stochastic fluctuation, we
further estimate the variances of the position $x$ and velocity
$v$ of the cantilever tip, which are defined by:
\begin{eqnarray*}
V_{x}=E\left(\left(x-E\left(x\right)\right)^2\right),\quad
V_{v}=E\left(\left(v-E(v)\right)^2\right).
\end{eqnarray*}
With additional calculations, the stationary values of the
variances $V_{x}$ and $V_{v}$ can be approximately estimated as
(see the analysis in Appendix~\ref{Derivations of the main results
in Sec.IV}):
\begin{equation}\label{Stationary variances under the white noise}
V_{x}^{\infty}=\frac{D_{\xi}v^*}{2F_{f_1}\left(k_c+k_x\right)},\quad
V_{v}^{\infty}=\frac{D_{\xi}v^*}{2F_{f_1}m}.
\end{equation}

The effectiveness of our proposal can be demonstrated via
numerical examples. Given the system parameters
\begin{eqnarray}\label{System parameters2}
&F_{f_0}=10\,\,{\rm nN},\quad F_{f_1}=1\,\,{\rm nN},\quad a=0.25\,\,{\rm nm},&\nonumber\\
&v_1=1\,\,{\rm
nm/s},\quad F_{f_2}=15\,\,{\rm nN},&\nonumber\\
&v^*=3\,\,{\rm nm/s}, \quad \sqrt{D_{\xi}}=0.1\,\,{\rm nN\cdot
s^{1/2}},\nonumber\\
&k_c=1\,\,{\rm N/m},\quad m=5\times10^{-11}\,\,{\rm
kg},\quad k_I=1\,\,{\rm N/\left(m\cdot s\right)},&\nonumber\\
&k_x=5\,\,{\rm N/m},\quad T=0.1\,\,{\rm s},\quad
\gamma_{ft}=1\,\,{\rm s}^{-1},&
\end{eqnarray}
we compare the motions of the cantilever tip driven by the
feedback control given in Eq.~(\ref{Real-time feedback control})
and by the open-loop control $u=v^* t$. Each curve is obtained by
averaging over $20$ sample (stochastic) trajectories. As shown in
Fig.~\ref{Plot of the motion of the sample pushed by feedback
control}, the stick-slip motion observed under open-loop control
(the green triangle curves) is greatly reduced by the proposed
feedback control (the blue solid curves). To compare the
stochastic fluctuations with the mean trajectories, we calculate
the square roots of the variances $\sigma_{x}=V_{x}^{1/2}$ and
$\sigma_{v}=V_{v}^{1/2}$. It can be obtained over $20$ stochastic
trajectories that $\sigma_{x}^{\infty}=0.06$ nm and
$\sigma_{v}^{\infty}=0.26$ nm/s, which are quite close to the
estimated values $\sigma_{x}^{\infty}=0.05$ nm,
$\sigma_{v}^{\infty}=0.38$ nm/s given by Eq.~(\ref{Stationary
variances under the white noise}) and are negligible compared with
the average motion ($E\left(x\right)$ is about tens of nm, and
$E\left(v\right)$ is about $3$ nm/s).

\begin{figure}[h]
\centerline{\includegraphics[bb=15 200 590 602,width=4 cm,
clip]{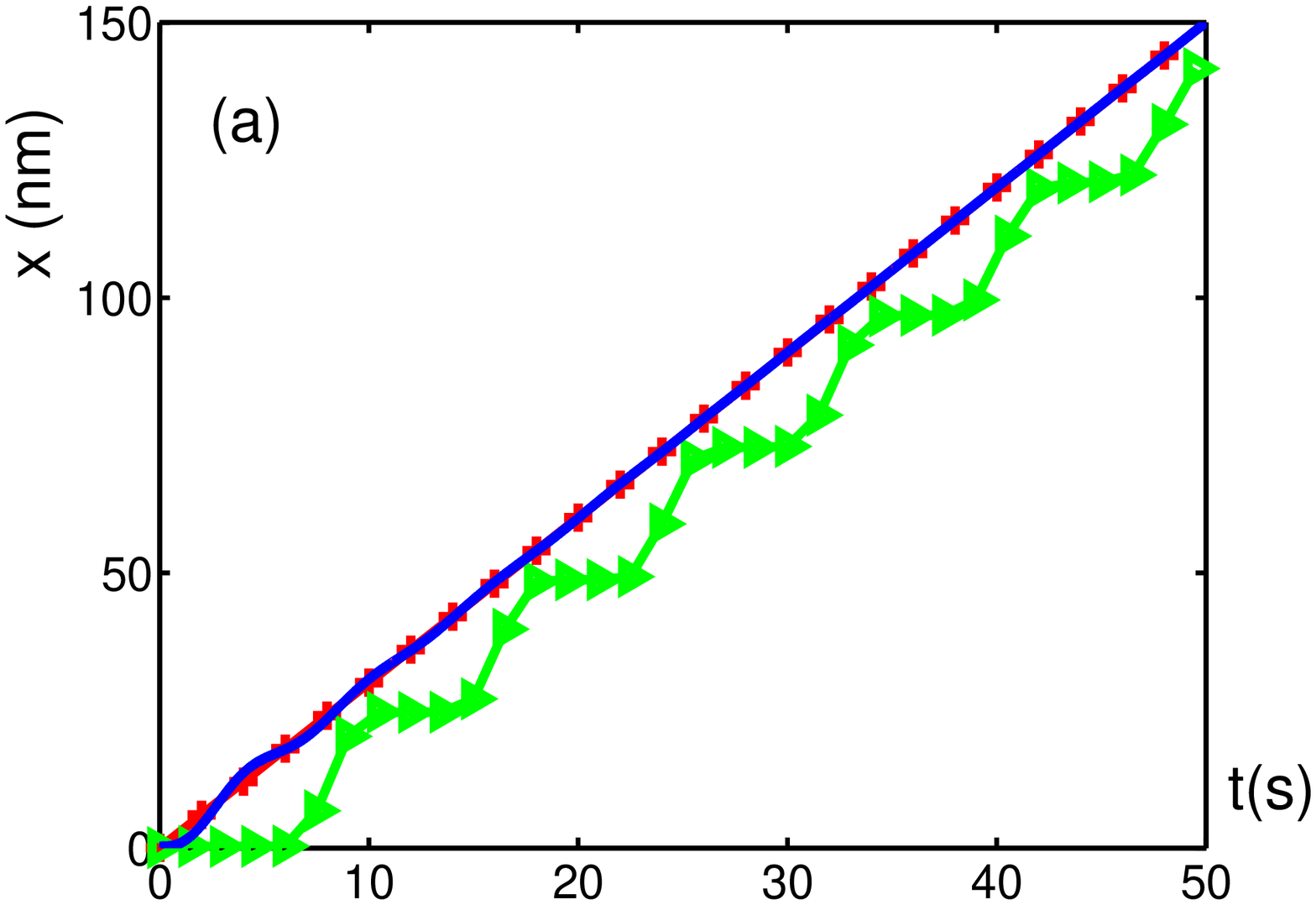}
\includegraphics[bb=26 195 587 601,width=4 cm, clip]{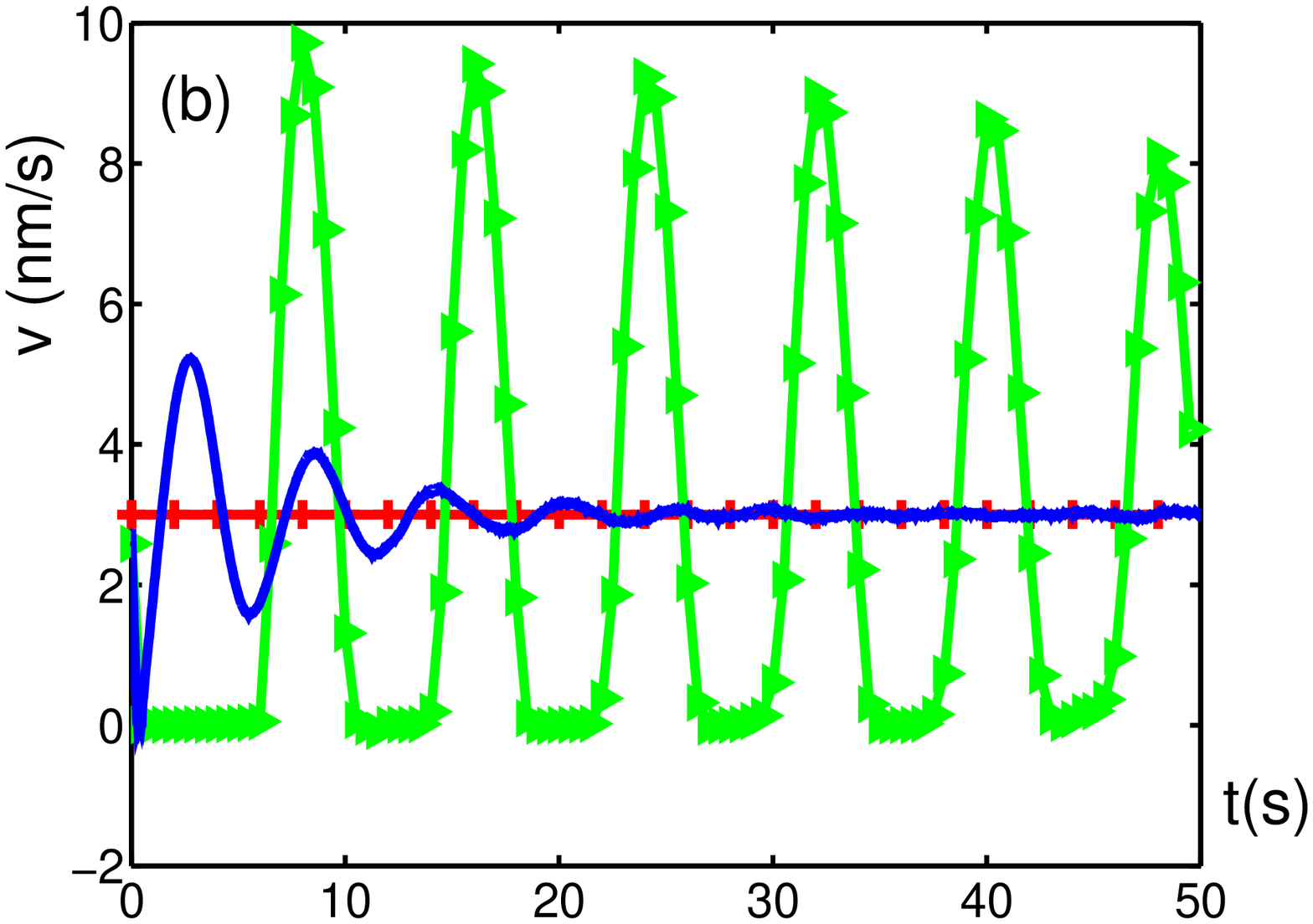}}
\centerline{\includegraphics[bb=15 200 590 602,width=4 cm,
clip]{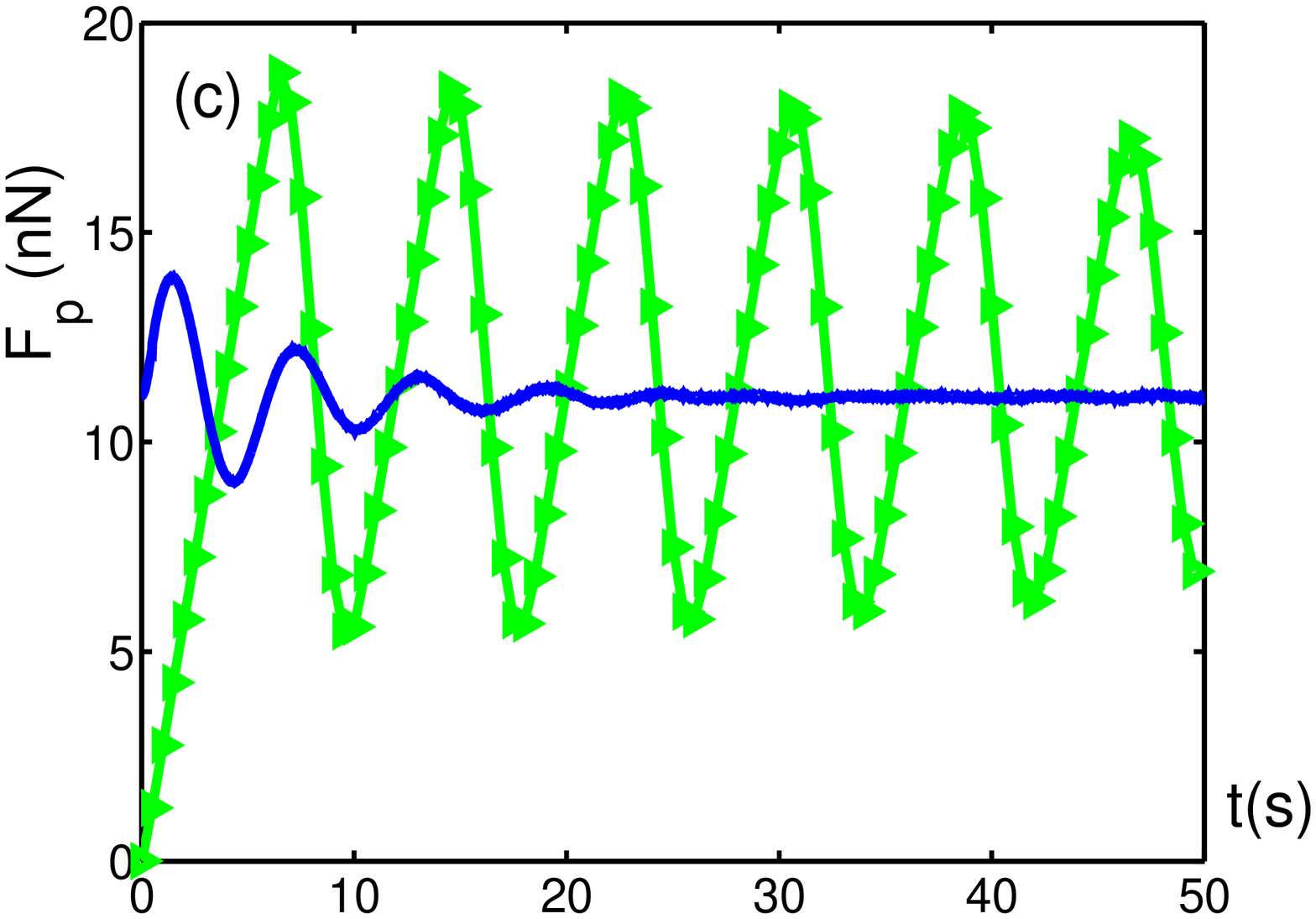}\includegraphics[bb=15 200 590 602,width=4 cm,
clip]{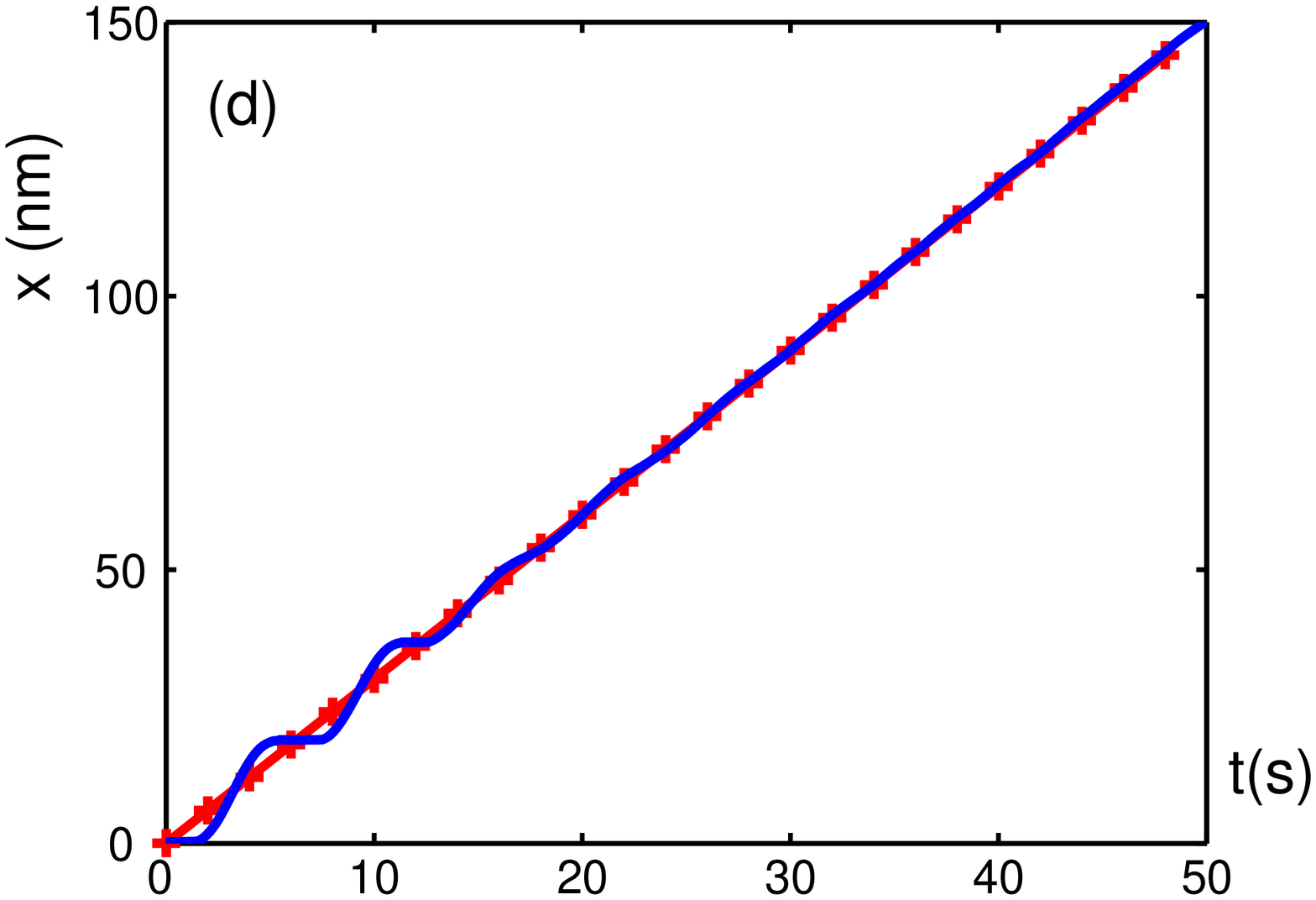}}\caption{\scriptsize (color online) Plots of the
motions of the cantilever tip with (a) the position $x$, (b) the
velocity $v$, and (c) the lateral force $F_p$ imposed by the
cantilever tip; (d) shows the trajectories in presence of
parameter uncertainties given in Eq.~(\ref{Uncertainties of the
system parameters}). The red curves with plus signs are the ideal
trajectories with $v=v^*=3$ nm/s; the green curves with triangle
signs represent the trajectories driven by the open-loop control
$u=v^* t$; and the blue solid curves denote the mean trajectories
under the feedback control given in Eq.~(\ref{Real-time feedback
control}). The mean trajectories in the plot are obtained over
$20$ stochastic trajectories.}\label{Plot of the motion of the
sample pushed by feedback control}
\end{figure}

\section{Robustness against parameter uncertainties}\label{s5}

The system parameters involved in Eq.~(\ref{System parameters2})
can be identified offline by, e.g., pre-designed nanofriction
experiments. In practical experiments, we also need to consider
the uncertainties in the system parameters which may deteriorate
the performances of the nano manipulation. For example, the
plastic deformation of the tip and the adhesion force between the
tip and the substrate can lead to small deviation $\Delta k_c$ of
the stiffness $k_c$ (typically $\Delta k_c/k_c<10\%$ in the
literature~\cite{Muller}). To reduce the effects of uncertainties,
we add an integral control term
$-k_I\int_0^t\left(\hat{x}-x^*\right)d\tau$ in Eq.~(\ref{Real-time
feedback control}) to reduce the static error induced by the
uncertainties.

Denote the additive uncertainties in the system parameters $k_c$,
$F_{f_0}$, $F_{f_1}$, $F_{f_2}$, $v_1$, and $a$ by $\Delta k_c$,
$\Delta F_{f_0}$, $\Delta F_{f_1}$, $\Delta F_{f_2}$, $\Delta
v_1$, $\Delta a$, and assume that there exists a phase offset
$\Delta\phi$ in the sine function in Eq.~(\ref{Flutuations in the
lateral force}). With the analysis given in
Appendix~\ref{Robustness analysis}, the static tracking error can
be controlled to zero, i.e.,
$\lim_{t\rightarrow\infty}\left(E\left(x\right)-x^*\right)=0$, if
the uncertainties of the system parameters are not too large to
satisfy
\begin{eqnarray}\label{Bounds of the uncertainties}
&\Delta_x=\left|\Delta k_c\right|+2\pi
\frac{F_{f_2}}{a^2}\left|\Delta
a\right|+\frac{2\pi}{a}\left|\Delta
F_{f_2}\right|<k_c+k_x,&\nonumber\\
&\left|\Delta F_{f_1}\right|<F_{f_1},&
\end{eqnarray}
and the gain of the integrator $k_I$ is chosen such that
\begin{equation}\label{Upper bound of kI}
k_I<\left(F_{f_1}-\left|\Delta
F_{f_1}\right|\right)\left(k_c+k_x-\Delta_x\right)/mv^*.
\end{equation}
It can be seen that the proportional and integral feedback terms
$k_x\left(\hat{x}-x^*\right)$,
$k_I\int_0^t\left(\hat{x}-x^*\right)d\tau$ in the control
(\ref{Real-time feedback control}) both contribute to the
robustness of our method: (i) the integral term is used to reduce
the static error; and (ii) the proportional term is used to
increase the robustness of our method about the parameter
uncertainty (the parameter regime given in Eqs.~(\ref{Bounds of
the uncertainties}) and (\ref{Upper bound of kI}) is enlarged when
we increase the control parameter $k_x$). Given the parameter
uncertainties:
\begin{eqnarray}\label{Uncertainties of the system parameters}
&\Delta F_{f_0}=1\,\,{\rm nN},\,\,\,\Delta F_{f_1}=0.1\,\,{\rm nN},\,\,\,\Delta a=0.01\,\,{\rm nm},&\nonumber\\
&\Delta F_{f_2}=0.1\,\,{\rm nN},\,\,\,\Delta k_c=0.1\,\,{\rm
N/m},&
\end{eqnarray}
Figure~\ref{Plot of the motion of the sample pushed by feedback
control}(d) shows that our method is still valid under the given
parameter uncertainties (the controlled trajectory, i.e., the blue
solid curve, matches very well with the ideal trajectory, i.e.,
the red curve with plus signs).

\section{Conclusion}\label{s6}

In summary, we propose a feedback control strategy to reduce the
stick-slip motion of the cantilever tip in a AFM-based nano
manipulation system. The feedback control is designed based on the
position estimation of the cantilever tip obtained by the force
sensing capacity of the AFM. Compared with open-loop control, our
proposal can greatly reduce the stick-slip motion of the
cantilever tip. Our method is robust against small uncertainties
in the system parameters, e.g., the stiffness of the cantilever of
the AFM, the lattice constant of the substrate, and the phase
offset in the surface potential of the substrate.

Future study will be focused on extending the method to more
practical cases. For example, as shown in our discussions, our
designed feedback control is valid only under small uncertainties.
More robust design should be developed for large uncertainties.
Since the uncertainties in the model of the system can be reduced
to a DC plus periodic disturbance, internal model principle is a
good choice of the control design. Additionally, our control
design can be naturally extended to the two-dimensional nano
manipulation systems. More interesting results, such as the
suppression of the S-shaped motion of the cantilever tip, are
hopeful to be observed for the two-dimensional case.

\begin{acknowledgments}
J. Zhang would like to thank Prof. Y.-X. Liu for helpful
discussions. J. Zhang and R. B. Wu are supported by the National
Natural Science Foundation of China under Grant Nos. 61174084,
61134008, 60904034. T.~J. Tarn would also like to acknowledge
partial support from the U. S. Army Research Office under Grant
W911NF-04-1-0386.
\end{acknowledgments}

\appendix
\section{Derivation of Eq.~(\ref{Exponential convergence of the position
estimation})}\label{Derivation of the exponential convergence of
the position estimation}

From Eqs.~(\ref{Measurement equation}) and (\ref{Estimations of
the position}), we have
\begin{eqnarray}
u-x&=&\left(F_p^m-\eta\left(t\right)\right)/k_c,\label{u-x}\\
u-\hat{x}&=&\hat{F}_p^m/k_c.\label{u-xhat}
\end{eqnarray}
It can be calculated from Eqs.~(\ref{Estimation of the measured
pushing force}) and (\ref{u-x}) that
\begin{eqnarray}\label{Exponential convergence of u-x}
&&\frac{1}{T}\int_{t-T}^t
e^{-\gamma_{ft}\left(t-\tau\right)}\left(u\left(\tau\right)-x\left(\tau\right)\right)d\tau\nonumber\\
&=&\frac{1}{k_c}\left(\hat{F}_p^m-\frac{1}{T}\int_{t-T}^t
e^{-\gamma_{ft}\left(t-\tau\right)}\eta\left(\tau\right) d\tau\right)\nonumber\\
&=&\frac{1}{k_c}\left(\hat{F}_p^m-\frac{1}{T}\int_0^T
e^{-\gamma_{ft}\tilde{\tau}}\eta\left(t-\tilde{\tau}\right) d\tilde{\tau}\right)\nonumber\\
&\approx&\frac{1}{k_c}\left(\hat{F}_p^m-\frac{1}{T}\int_0^T
\eta\left(t-\tilde{\tau}\right) d\tilde{\tau}\right)\nonumber\\
&=&\frac{1}{k_c}\left(\hat{F}_p^m-\frac{1}{T}\int_{t-T}^t
\eta\left(\tau\right)
d\tau\right)\nonumber\\
&\approx&\frac{1}{k_c}\hat{F}_p^m=u-\hat{x}.
\end{eqnarray}
Here, we have used the condition given in Eq.~(\ref{Filtering
condition}), i.e., $T\ll1/\gamma_{ft}$, to obtain
$e^{-\gamma_{ft}t}\approx 1$ when $t\in\left[0,T\right]$ and the
ergodic property~\cite{Peterse} of the white noise
$\eta\left(t\right)$ to replace the time average of
$\eta\left(t\right)$ by its ensemble average, i.e.,
\begin{eqnarray*}
\frac{1}{T}\int_{t-T}^t\eta\left(\eta\right)d\tau\approx
E\left(\eta\left(t\right)\right)=0.
\end{eqnarray*}
Let us set
\begin{eqnarray*}
h\left(t\right)&=&u\left(t\right)-x\left(t\right),\\
\hat{h}\left(t\right)&=&u\left(t\right)-\hat{x}\left(t\right),
\end{eqnarray*}
then from Eq.~(\ref{Exponential convergence of u-x}), i.e.,
\begin{eqnarray*}
\hat{h}\left(t\right)=\frac{1}{T}\int_{t-T}^t
e^{-\gamma_{ft}\left(t-\tau\right)}h\left(\tau\right)d\tau,
\end{eqnarray*}
we have
\begin{equation}\label{Differential form of h}
\frac{d}{dt}\hat{h}\left(t\right)=\frac{1}{T}\left(h\left(t\right)-e^{-\gamma_{ft}T}h\left(t-T\right)\right)-\gamma_{ft}h\left(t\right).
\end{equation}
We assume that
$g\left(T\right)=e^{\gamma_{ft}T}h\left(t+T\right)$, then we have
\begin{eqnarray*}
\frac{1}{T}\left(h\left(t\right)-e^{-\gamma_{ft}T}h\left(t-T\right)\right)=\frac{1}{T}\left(g\left(0\right)-g\left(-T\right)\right),
\end{eqnarray*}
which can be replaced by
\begin{eqnarray*}
g'\left(T\right)\left|_{T=0}\right.=\frac{d}{dt}h\left(t\right)+\gamma_{ft}h\left(t\right)
\end{eqnarray*}
under the condition $T\ll1/\gamma_{ft}$. Thus,
Eq.~(\ref{Differential form of h}) can be reexpressed as:
\begin{equation}\label{Asymptotical convergence of h}
\frac{d}{dt}\left(\hat{h}-h\right)=-\gamma_{ft}\left(\hat{h}-h\right).
\end{equation}
From the definition of $h$ and $\hat{h}$, we have
\begin{equation}\label{Asmptotical convergence of the position of the cantilever tip}
\frac{d}{dt}\left(\hat{x}-x\right)=-\gamma_{ft}\left(\hat{x}-x\right),
\end{equation}
which means that $\hat{x}-x\rightarrow 0$ when
$t\gg1/\gamma_{ft}$, i.e., the estimated position of the
cantilever tip $\hat{x}$ tracks the actual position $x$ in the
long time limit.

\section{Derivation of Eqs.~(\ref{Position tracking}) and (\ref{Stationary variances under the white noise})}\label{Derivations of the main results in Sec.IV}         

By substituting the feedback control law given in
Eq.~(\ref{Real-time feedback control}) into Eq.~(\ref{Dynamical
equations of the sample without defection}), we have
\begin{eqnarray}\label{Stochastic dynamic and estimation equations}
dx&=&v dt,\nonumber\\
mdv&=&\left[-k_c\left(x-x^*\right)-k_x\left(\hat{x}-x^*\right)\right.\nonumber\\
&&+F_{f_2}\sin\left(2\pi\frac{\hat{x}}{a}\right)-F_{f_2}\sin\left(2\pi\frac{x}{a}\right)\nonumber\\
&&\left.-F_{f_1}\left({\rm ln}\frac{v}{v_1}-{\rm
ln}\frac{v^*}{v_1}\right)-k_I q\right]dt+\sqrt{D_{\xi}}dW,\nonumber\\
d\delta\hat{x}&=&-\gamma_{ft}\delta\hat{x} dt,\,\,\,d
q=\left(\hat{x}-x^*\right)dt,
\end{eqnarray}
where $\delta\hat{x}=\hat{x}-x$ and
$q=\int_0^t\left(\hat{x}-x^*\right)d\tau$.

By taking the ensemble average and denoting $\delta
\bar{x}=E\left(x\right)-x^*$, $\delta
\bar{v}=E\left(v\right)-v^*$,
$\delta\bar{\hat{x}}=E\left(\delta\hat{x}\right)$, and
$\bar{q}=E\left(q\right)$, we can expand the equation to the
second-order quadratures to obtain:
\begin{eqnarray}\label{Ensemble average equation}
\delta\dot{\bar{x}}&=&\delta\bar{v},\nonumber\\
m\delta\dot{\bar{v}}&=&-\left(k_c+k_x\right)\delta\bar{x}+F_{f_2}\frac{2\pi}{a}\cos\left(2\pi\frac{x^*}{a}\right)\delta\bar{\hat{x}}\nonumber\\
&&-k_x\delta\bar{\hat{x}}-\frac{F_{f_1}}{v^*}\delta\bar{v}+F_{f_2}\frac{8\pi^3}{a^3}V_{x}\cos\left(2\pi\frac{x^*}{a}\right)\delta\bar{\hat{x}}\nonumber\\
&&+F_{f_1}\frac{V_{v}}{2v^{*2}}+O\left(\frac{\delta\bar{x}}{a}\frac{\delta\bar{\hat{x}}}{a}\right)+O\left(\frac{\delta\bar{v}^2}{v^{*2}}\right)\nonumber\\
&&+O\left(\frac{V_{x}^2}{a^4}\frac{\delta\bar{\hat{x}}}{a}\right)+O\left(\frac{V_{v}^2}{v^{*4}}\right)-k_I\bar{q},\nonumber\\
\delta\dot{\bar{\hat{x}}}&=&-\gamma_{ft}\delta\bar{\hat{x}},\,\,\,\dot{\bar{q}}=\delta\bar{x}+\delta\bar{\hat{x}},
\end{eqnarray}
where $V_{x}=E\left(x-E\left(x\right)\right)^2$ and
$V_{v}=E\left(v-\left(v\right)\right)^2$ are the variances of $x$
and $v$; and $O\left(\cdot\right)$ denotes the higher-order terms.
Since the nonlinear equation (\ref{Stochastic dynamic and
estimation equations}) is linearized near the origin
$\left(0,0,0,0\right)^T$ in Eq.~(\ref{Ensemble average equation}),
we have introduced the Gaussian assumption to omit higher-order
quadratures of $x$ and $v$. We can further omit the terms related
to $V_{x}/a^2$ and $V_{v}/v^{*2}$ in Eq.~(\ref{Ensemble average
equation}). In fact, as shown in Eq.~(\ref{Stationary variances
under the white noise}), we have $V_{x}^{\infty}\ll a^2$,
$V_{v}^{\infty}\ll v^{*2}$ under the assumption (\ref{Tiny
fluctuation condition}). Thus, we have $V_{x}\ll a^2$ and
$V_{v}\ll v^{*2}$ for sufficiently long time. Since we are just
interested in the stationary behaviors of the system dynamics, we
can omit the terms related to $V_{x}/a^2$ and $V_{v}/v^{*\,2}$.

With the above analysis, we can obtain the linearization equation
of Eq.~(\ref{Ensemble average equation}) in the neighborhood of $
\left(\delta\bar{x},\delta\bar{v},\delta\bar{\hat{x}},\bar{q}\right)^T=\left(0,0,0,0\right)^T$.
It can be easily verified that the characteristic equation of the
coefficient matrix of the linearization equation is
\begin{eqnarray*}
\left(s+\gamma_{ft}\right)\left(s^3+\frac{F_{f_1}}{mv^*}s^2+\frac{\left(k_c+k_x\right)}{m}s+\frac{k_I}{m}\right)=0.
\end{eqnarray*}
Since $k_I<F_{f_1}\left(k_c+k_x\right)/mv^*$ from Eq.~(\ref{kI
without uncertainties}), the real parts of the eigenvalues of the
above equation are all negative. It means that the linearization
equation is exponentially stable, and thus the original nonlinear
equation (\ref{Ensemble average equation}) is asymptotically
stable at the origin, which leads to the fact that
$\lim_{t\rightarrow\infty}\left(E\left(x\right)-x^*\right)=0$.

To approximately estimate the stationary variances, we replace
$\hat{x}$ by $x$, linearize Eq.~(\ref{Stochastic dynamic and
estimation equations}), and omit higher-order correlation terms.
It can be verified that $V_{x}$, $V_{v}$, and $C_{x v}$, i.e., the
covariance between $x$ and $v$, satisfy the following equation
\begin{eqnarray}\label{Equation of the variances and covariance}
\left(%
\begin{array}{c}
  \dot{V}_{x} \\
  \dot{V}_{v} \\
  \dot{C}_{x v} \\
\end{array}%
\right)&=&\left(%
\begin{array}{ccc}
  0 & 0 & 2 \\
  0 & -\frac{2F_{f_1}}{mv^*} & -\frac{2\left(k_c+k_x\right)}{m} \\
  -\frac{k_c+k_x}{m} & 1 & -\frac{F_{f_1}}{mv^*} \\
\end{array}%
\right)\nonumber\\
&&\left(%
\begin{array}{c}
  V_{x} \\
  V_{v} \\
  C_{x v} \\
\end{array}%
\right)+\left(%
\begin{array}{c}
  0 \\
  \frac{D_{\xi}}{m^2} \\
  0 \\
\end{array}%
\right).
\end{eqnarray}
To consider the stationary variances, the omission of the
higher-order correlation terms to obtain Eq.~(\ref{Equation of the
variances and covariance}) is reasonable, because
$x-\hat{x}\rightarrow 0$ when $t\rightarrow\infty$, and
higher-order correlation terms are small compared with $V_x$,
$V_v$, and $C_{xv}$ under the weak noise assumption (\ref{Tiny
fluctuation condition}). From Eq.~(\ref{Equation of the variances
and covariance}), we can calculate the stationary variances of $x$
and $v$ as:
\begin{eqnarray*}
V_{x}^{\infty}=\frac{D_{\xi}v^*}{2F_{f_1}\left(k_c+k_x\right)},\quad
V_{v}^{\infty}=\frac{D_{\xi}v^*}{2F_{f_1}m}.
\end{eqnarray*}

\section{Robustness analysis of our method}\label{Robustness analysis}

Let us replace the system parameters $k_c$, $F_{f_0}$, $F_{f_1}$,
$F_{f_2}$, $v_1$, and $a$ in Eq.~(\ref{Ensemble average equation})
by $k_c+\Delta k_c$, $F_{f_0}+\Delta F_{f_0}$, $F_{f_1}+\Delta
F_{f_1}$, $F_{f_2}+\Delta F_{f_2}$, $v_1+\Delta v_1$, and
$a+\Delta a$, and consider the phase offset $\Delta \phi$. We can
expand the equation to the linear terms of the uncertainties
$\Delta k_c$, $\Delta F_{f_0}$, $\Delta F_{f_1}$, $\Delta
F_{f_2}$, $\Delta v_1$, $\Delta \phi$, and $\Delta a$. By
neglecting the higher-order nonlinear terms, we can obtain
\begin{eqnarray}\label{Linearization equation with uncertainty}
\delta\dot{\bar{x}}&=&\delta\bar{v},\nonumber\\
m\delta\dot{\bar{v}}&=&-w_x\delta\bar{x}-w_v\delta\bar{v}-\left(k_x-F_{f_2}\frac{2\pi}{a}\cos\frac{x^*}{a}\right)\delta\bar{\hat{x}}\nonumber\\
&&-k_I\bar{q}+w_0,\nonumber\\
\delta\dot{\bar{\hat{x}}}&=&-\gamma_{ft}\delta\bar{\hat{x}},\,\,\,\dot{\bar{q}}=\delta\bar{x}+\delta\bar{\hat{x}},
\end{eqnarray}
where
\begin{eqnarray*}
w_x&=&-\frac{2\pi}{a}\cos\left(2\pi\frac{x^*}{a}\right)\left(F_{f_2}\frac{\Delta
a}{a}-\Delta F_{f_2}\right)\\
&&+\Delta k_c+k_c+k_x,\\
w_v&=&\left(F_{f_1}+\Delta F_{f_1}\right)/v^*,\\
w_0&=&\Delta F_{f_0}+{\rm ln}\left(\frac{v^*}{v_1}\right)\Delta
F_{f_1}+\frac{F_{f_1}}{v_1}\Delta
v_1\\
&&-\sin\left(2\pi\frac{x^*}{a}\right)\Delta
F_{f_2}-F_{f_2}\cos\left(2\pi\frac{x^*}{a}\right)\Delta \phi.
\end{eqnarray*}
The characteristic equation of the linear matrix of
Eq.~(\ref{Linearization equation with uncertainty}) can be
expressed as:
\begin{eqnarray*}
\left(s+\gamma_{ft}\right)\left(s^3+w_v s^2+w_x s+k_I\right)=0.
\end{eqnarray*}

It can be checked from Eqs.~(\ref{Bounds of the uncertainties})
and (\ref{Upper bound of kI}) that the real parts of the
eigenvalues of the above equation are all negative. It means that
there exists a stationary solution of the linearlization equation
of Eq.~(\ref{Ensemble average equation}), so does the original
equation~(\ref{Ensemble average equation}). Thus, we have
$\dot{\bar{q}}=\delta\bar{x}+\delta\bar{\hat{x}}\rightarrow 0$
when $t\rightarrow \infty$. Furthermore, from
$\delta\dot{\bar{\hat{x}}}=-\gamma_{ft}\delta\bar{\hat{x}}$, it
can be verified that $\delta\bar{\hat{x}}\rightarrow 0$. Thus, we
have $\lim_{t\rightarrow
\infty}\delta\bar{x}=\lim_{t\rightarrow\infty}\left(E\left(x\right)-x^*\right)=0$.


\end{document}